\documentclass[a4paper,12pt]{article}
\usepackage[utf8]{inputenc}
\usepackage[english]{babel}
\usepackage[round]{natbib}
\usepackage[margin=2.5cm]{geometry}
\usepackage{amsmath}
\usepackage{mathrsfs}
\usepackage{longtable}
\usepackage{booktabs}
\usepackage[referable]{threeparttablex}
\usepackage[font=small,labelfont=bf,singlelinecheck=on]{caption}
\usepackage{scalefnt}
\usepackage{booktabs}
\usepackage[referable]{threeparttablex}
\usepackage{graphicx}
\usepackage{hyperref}
\usepackage{setspace}
\usepackage{authblk}
\usepackage{fancyhdr}

\hypersetup{
    colorlinks=true,
    urlcolor=blue,
    linkcolor=blue,
    citecolor=blue}

\title{COVID-19 and Global Economic Growth: Policy Simulations with a Pandemic-Enabled Neoclassical Growth Model}

\author[1]{Ian M. Trotter\thanks{Corresponding author. E-mail: \href{mailto:ian.trotter@ufv.br}{\texttt{ian.trotter@ufv.br}}}}
\author[1]{Luís A. C. Schmidt}
\author[1]{Bruno C. M. Pinto}
\author[2]{Andrezza L. Batista}
\author[2]{Jéssica L. V. Pellenz}
\author[2]{Maritza Rosales}
\author[2]{Aline Rodrigues}
\author[2]{Attawan G. S. Suela}
\author[2]{Loredany C. C. Rodrigues}

\affil[1]{\normalsize Institute of Public Policy and Sustainable Development, Universidade Federal de Viçosa}
\affil[2]{\normalsize Department of Agricultural Economics, Universidade Federal de Viçosa}

\date{\textbf{Working Paper} -- June 2020}

\begin{document}

\maketitle

\begin{abstract}
    \noindent
    During the COVID-19 pandemic of 2019/2020, authorities have used temporary \textit{ad-hoc} policy measures, such as lockdowns and mass quarantines, to slow its transmission. However, the consequences of widespread use of these unprecedented measures are poorly understood. To contribute to the understanding of the economic and human consequences of such policy measures, we therefore construct a mathematical model of an economy under the impact of a pandemic, select parameter values to represent the global economy under the impact of COVID-19, and perform numerical experiments by simulating a large number of possible policy responses. By varying the starting date of the policy intervention in the simulated scenarios, we find that the most effective policy intervention occurs around the time when the number of active infections is growing at its highest rate -- that is, the results suggest that the most severe measures should only be implemented when the disease is sufficiently spread. The intensity of the intervention, above a certain threshold, does not appear to have a great impact on the outcomes in our simulations, due to the strongly concave relationship that we identify between production shortfall and infection rate reductions. Our experiments further suggest that the intervention should last until after the peak established by the reduced infection rate, which implies that stricter policies should last longer. The model and its implementation, along with the general insights from our policy experiments, may help policymakers design effective emergency policy responses in the face of a serious pandemic, and contribute to our understanding of the relationship between the economic growth and the spread of infectious diseases.
\end{abstract}

\section{Introduction}
Pandemics have caused death and destruction several times throughout human history, and have caused large and lasting impacts on society: for example the Black Death in 14\textsuperscript{th} century Europe \citep{herlihy_black_1997}, the Spanish flu in 1918-1920 \citep{johnson_updating_2002}, HIV in the 1980s \citep{pope_transmission_2003}, and H1N1 in 2009 \citep{trifonov_origin_2009}. In 2019 and 2020, policymakers have struggled to design effective policy responses to the COVID-19 pandemic, as authorities have resorted to unprecedented and widespread use of temporary emergency measures such as lockdowns, mass quarantines, and other ``social distancing'' measures, despite that the human and economic consequences of these \textit{ad-hoc} interventions are poorly understood. It is clear that more research is needed on effective policy intervention during pandemics, and how to mitigate its human and economic impacts \citep{bauch_covid-19_2020}.

Therefore, we incorporate a model of disease transmission into a model of economic growth, considering that policymakers can implement temporary policies that simultaneously slow the spread of the disease and lower the economic output. We then select model parameters so as to represent the global economy under the impact of COVID-19, and perform numerical simulations of various policies to provide insights into the trade-offs between short- and long-term human and economic outcomes. The policy simulations will help policymakers understand how altering the starting time, intensity, and duration of the policy intervention can impact the outcomes, and contribute to the understanding of how to design effective policies to confront rapidly spreading pandemics.

A growing body of literature has been devoted to studying the economic impacts of social distancing measures and the design of emergency policies during pandemics. For instance, \cite{eichenbaum_macroeconomics_2020} constructed a mathematical model and ran numerical experiments that resulted in important insights for policymakers during a pandemic, although the study abstracts from forces that affect the long-term economic development. \cite{andersson_optimal_2020} use a similar framework to study policy responses and the trade-off between output and health during a pandemic, operating with a welfare function that only includes the period of the epidemic, and will also not account for lasting impacts of the pandemic on the economy. Our study is similar to these studies, although our model includes capital stock and population growth, which enables us to look at possible long-term effects beyond the duration of the pandemic. \cite{guan_global_2020} consider the global economic impacts of the COVID-19 pandemic in a model with multiple countries and productive sectors, suggesting that the economic impacts propagate through supply chains, and that earlier, stricter, and shorter lockdowns will minimise the economic damages. \cite{acemoglu_optimal_2020} include a simplified evaluation of economic loss into an SIR-type epidemiological model that distinguishes between age groups, and find that lockdown policies that are tailored to the different age groups are more efficient. \cite{la_torre_optimal_2020} model the social costs of an epidemic, in which a social planner must choose to allocate tax revenues between prevention and treatment, showing that the optimal allocation depends on the infectivity rate of the epidemic. \cite{alvarez_simple_2020} studied the optimal lockdown policy in a linear economy, where each fatality incurs a cost.

However, even before the onset of the COVID-19 pandemic, several studies had focused on the determinants and incentives of social distancing during epidemics \citep{fenichel_economic_2013, perrings_merging_2014, kleczkowski_spontaneous_2015, toxvaerd_equilibrium_2020, quaas_social_2020}. The models in these studies mainly apply economic insights to improve epidemiological models by examining and modelling individual contact decisions.

On the relationship between epidemics and economic growth, \cite{herlihy_black_1997} and \cite{hansen_malthus_2002} argue that the increased mortality caused by the Black Death resulted in great economic damage by decreasing labour supply, which induced the substitution of labour for capital and triggered an economic modernisation that eventually lead to greater economic growth. \cite{delfino_positive_2000}, however, argue that there are causal links in both directions in the interaction between the economy and disease transmission, and proposed a model that combines disease transmission into a model of economic growth, although the model is not used to explore temporary policy interventions during specific rapidly-spreading pandemics. \cite{bonds_poverty_2010}, \cite{ngonghala_poverty_2014} and \cite{ngonghala_general_2017} show that poverty traps\footnote{That is, in this particular case, that a greater disease burden leads to more poverty, and more poverty leads to a greater disease burden, so as to create a self-perpetuating effect.} frequently arise when combining models of infectious diseases and of economic development, and that these may help explain the differences in economic development between countries. \cite{goenka_infectious_2010}, \cite{goenka_infectious_2014} and \cite{goenka_infectious_2019} also integrate disease transmission directly into an economic growth model, and allow investments in health -- building \textit{health capital} -- to affect the epidemiological parameters. Optimal investments in health and the accumulation of health capital, however, are different from designing temporary policies during a pandemic. Most of these studies are based on disease models that allow the individual to contract the disease multiple times, which is consistent with some diseases that are common in the developing world, such as malaria or dengue fever, but may not be applicable to COVID-19. Furthermore, most of these models do not include disease-related mortality, one of the main channels through which serious pandemics affect the economy \citep{hansen_malthus_2002}. These studies, however, show that integrating disease transmission in economic growth models can result in multiple steady states, and we are therefore careful to choose an approach that is not sensitive to this: we use our model to run numerical experiments, solving the numerical optimisation problem using backwards induction.

The macroeconomic impact of the HIV/AIDS epidemic has also received much research attention, altough many studies assume the disease transmission is exogenous \citep{cuddington_modeling_1993, cuddington_further_1993, cuddington_assessing_1994, haacker_modeling_2002, arndt_hiv-aids_2003, cuesta_how_2010}. \cite{azomahou_hivaids_2016} allow mortality rate to depend on health expenditure, but the disease transmission still remains exogenous to their model. \cite{bell_macroeconomics_2009} studied government investments in health and education in an overlapping generations model of HIV/AIDS in Africa. However, the HIV/AIDS epidemic and the COVID-19 pandemic are so different, that we do not expect the insights from these studies to transfer directly to the COVID-19 pandemic.

The COVID-19 pandemic has shown that many authorities are prepared to implement strict and dramatic emergency policies at short notice in order to slow down the spread of a serious pandemic. However, our understanding of the economic and human consequences of such measures is still incipient. At the same time, implementing such policies is delicate, and, if done improperly, authorities could damage the economy whilst still failing to lower the transmission rate of the disease. There is therefore an urgent need for research that develops guidelines for the use of these emergency measures, and to help policymakers understand their impacts and consequences over time.

We contribute to the study of the efficiency, impacts, and consequences of temporary emergency measures during a pandemic by incorporating a policy parameter into a model that integrates disease transmission dynamics and economic growth. Our model provides a theoretical framework for understanding the impact of emergency policies on the trajectory of the pandemic as well as the main economic variables, in light of their mutual interactions. To gain a deeper understanding of the emergency policies, we select parameter values that are consistent with the global economy under the impact of COVID-19, and numerically simulate a large number of scenarios for possible emergency policies. Using this simulation-based approach, we investigate the impact of altering the starting date of policy intervention, the intensity of the policy intervention, and the duration of the policy intervention. Altering the simulated policies along these three dimensions provides insights into the impact of the emergency measures on the trajectory of the pandemic and the development of the main economic variables. These insights could help policymakers design effective emergency responses to pandemics.

Our work differs from earlier studies in some important aspects. Principally, the period of interest in our study extends beyond the duration of the pandemic. Therefore, our model includes the dynamics of capital accumulation, population growth and pandemic deaths, which have not been jointly considered in previous studies. These components are important to study the impact of the pandemic on economic growth beyond the short- and medium term. Other novel aspects of our model include a relationship between the reduction in economic output and the reduction in the infection rate in the short run, a mortality rate that depends on the infection rate, and explicitly modelled excess costs of hospital admissions due to the pandemic. Although some of these features are included in previous models, they have not yet been combined into a single comprehensive framework. In addition, our study contains some early estimates of the economic impacts of COVID-19 on Europe's five largest economies, constructed by analysing changes in real-time and high-frequency data on electricity demand. We also make our data and custom computer code freely available, which will hopefully be useful to the research community and contribute to future developments in the area.

In addition to this introduction, this paper consists of three sections. The following section presents a mathematical framework that incorporates a model of disease transmission into a model of economic growth, shows how model parameter values were chosen to fit the model to the global economy during the COVID-19 pandemic, and details the numerical experiments that were performed. In the third section, we present, interpret, and discuss the results of the numerical experiments and their implications. In the final section, we summarise the main findings of the study.

\section{Pandemic in an Economic Growth Model}
Here we detail the integration of an epidemiological model into a neoclassical model of economic growth -- known as one of the ``workhorses'' of modern macroeconomics \citep{acemoglu_introduction_2011}. We first modify the SIR (Susceptible-Infected-Recovered) model of the spread of an infection, pioneered by \cite{kermack_contribution_1927}, then incorporate it into the a model setup similar to the classical Ramsey-Cass-Koopmans model in discrete time. We then explain how we select functional forms and parameters to represent the global economy and the global spread of COVID-19, before outlining a set of numerical experiments designed to give insight into the economic and epidemiological impacts of varying the starting time, intensity and duration of the policy interventions.

\subsection{Incorporating Pandemic Dynamics into a Neoclassical Model of Economic Growth}
Here we combine an epidemiological model with a model of economic growth, concentrating on three main bridges between the models. First, we assume that the spread of a pandemic reduces the labour force, since infected or deceased individuals will not work, and this reduces economic output. Second, we assume that society incurs additional direct costs, for instance due to the hospitalisation of infected individuals, and these costs must be covered with output that would otherwise have been consumed or invested. Third, we assume that governments may, through policy, simultaneously impact both the spread of the pandemic and the efficiency of economic production. These interactions between the spread of the pandemic and economic growth are the main focus of our model, which jointly represents the dynamics of the spread of the pandemic and the dynamics of economic growth.

The SIR model \citep{kermack_contribution_1927, brauer_mathematical_2012} is a simple Markov model of how an infection spreads in a population over time. This model divides a population ($N$) into three categories: Susceptible ($S$), Infected ($I$), and Recovered ($R$). In each period, the number of susceptible individuals who become infected is a product of the susceptible population, the number of individuals who are already infected, and an infection rate $b$. A given proportion of the infected individuals ($r$) also recover in each period.

To incorporate the SIR model into a model of economic growth, we make two adaptations to the basic SIR model. First, we introduce a distinction between recovered individuals ($R$) and deceased individuals ($D$), since recovered individuals will re-enter the labour force whereas deceased individuals will not: each period, infected individuals will recover at a rate $r$ and pass away at a rate $m$. Second, instead of considering the population to be of a fixed size, we allow the population to grow over time. Population growth is usually negligible at the timescale of interest for models of epidemics or pandemics, but it is significant in the timescales of economic growth. Therefore, we introduce a logistic model for population growth, and new individuals will be added to the number of susceptible individuals each period. Using two parameters, $a_1$ and $a_2$, to describe the population growth, we can describe the spread of the pandemic in the population as follows:
\begin{align}
    \label{eq:popgrowth}
    N_{t+1} &= a_1 N_t + a_2 N_t^2 - m I_t \\
    \label{eq:susceptible}
    S_{t+1} &= S_t + (a_1-1)N_t + a_2 N_t^2 - b S_t I_t \\
    \label{eq:infected}
    I_{t+1} &= I_t + b S_t I_t - r I_t - m I_t \\
    \label{eq:recovere}
    R_{t+1} &= R_t + r I_t \\
    \label{eq:deceased}
    D_{t+1} &= D_t + m I_t.
\end{align}
Many variations of the basic SIR model already exist, and it would be possible to incorporate more complex dynamics into the epidemiological model. However, this model will be sufficient for our current purposes.

The model of economic growth assumes that a representative household chooses what quantity of economic output ($Y$) to consume ($C$) or save (invest) each period in order to maximise an infinite sum of discounted utility, represented by a logarithmic utility function. Output is produced by combining labour and capital ($K$) using a technology represented by a Cobb-Douglas production function with constant returns to scale and total factor productivity $A_t$. However, we allow pandemic policy, represented by $p$, to reduce the total output, and furthermore assume that only susceptible and recovered individuals are included in the labour force:
\begin{gather}
    Y_t = (1-p) A_t K_t^\alpha (S_t+R_t)^{1-\alpha},
\end{gather}
in which $\alpha$ represents the output elasticity of capital, and total factor productivity $A_t$ grows at a constant rate $g$:
\begin{gather}
    A_{t+1} = (1+g)A_t.
\end{gather}
Assuming that physical capital ($K_t$) depreciates at a rate of $\delta$ from one period to the next and that the pandemic causes direct costs $H(\cdot)$ to society, the capital stock accumulates according to the following transition equation:
\begin{gather}
    \label{eq:capitalaccumulation}
    K_{t+1} = (1-\delta) K_t - C_t - H(\cdot).
\end{gather}

Given a utility discount factor $\beta$, we assume that a benevolent social planner chooses an infinite stream of consumption $\{ C_t \}_{t=0}^\infty$ to optimise the discounted sum of logarithmic utility, solving the following maximisation problem:
\begin{gather*}
    \max_{\{ C_t \}_{t=0}^\infty } \sum_{t=0}^\infty \beta^t N_t \ln{\left( \frac{C_t}{N_t} \right) },
\end{gather*}
while respecting the restrictions represented by equations \eqref{eq:popgrowth} through \eqref{eq:capitalaccumulation}. Since our period of interest is actually finite, this maximisation problem can be solved numerically by backwards induction, provided that the terminal period is chosen so far in the future that it will not interfere with the period of interest\footnote{The implementation of the model, \emph{Macroeconomic-Epidemiological Emergency Policy Intervention Simulator}, is available at \url{https://github.com/iantrotter/ME3PI-SIM}.}.

\subsection{Representing the Global Economy and COVID-19}
The model presented in the previous subsection relies on parameters for population dynamics, the spread of a pandemic, production of economic output, and the accumulation of physical capital. In order to perform computational experiments, we need to determine realistic numerical values for these, as well as initial values for the state variables.

One fundamental issue that we must address is that the study of epidemics and economic growth usually consider different timescales: whereas the spread of an epidemic is usually analysed at a daily or weekly timescale, economic growth is usually studied at an annual, or even a decennial, timescale. To reconcile these differences, we choose a daily timescale for our model. A daily timescale is suitable for studying the spread of a pandemic, since pandemics spread rapidly and their health effects pass almost entirely within a short timeframe. However, daily resolution is an unusual choice for a model of economic growth, as capital accumulation, technological progress, and population growth are almost negligible from one day to the next. During a pandemic, however, daily movements of individuals in and out of the labour force could have a large impact on economic production, and indirectly on the accumulation of capital -- and in order to adequately capture these effects, we choose a daily resolution for the model. Parameter values for both the parameters belonging to the economic components, as well as the epidemiological components, are therefore chosen to represent a daily timescale.

\begin{figure}
    \centering
    \includegraphics[width=\textwidth]{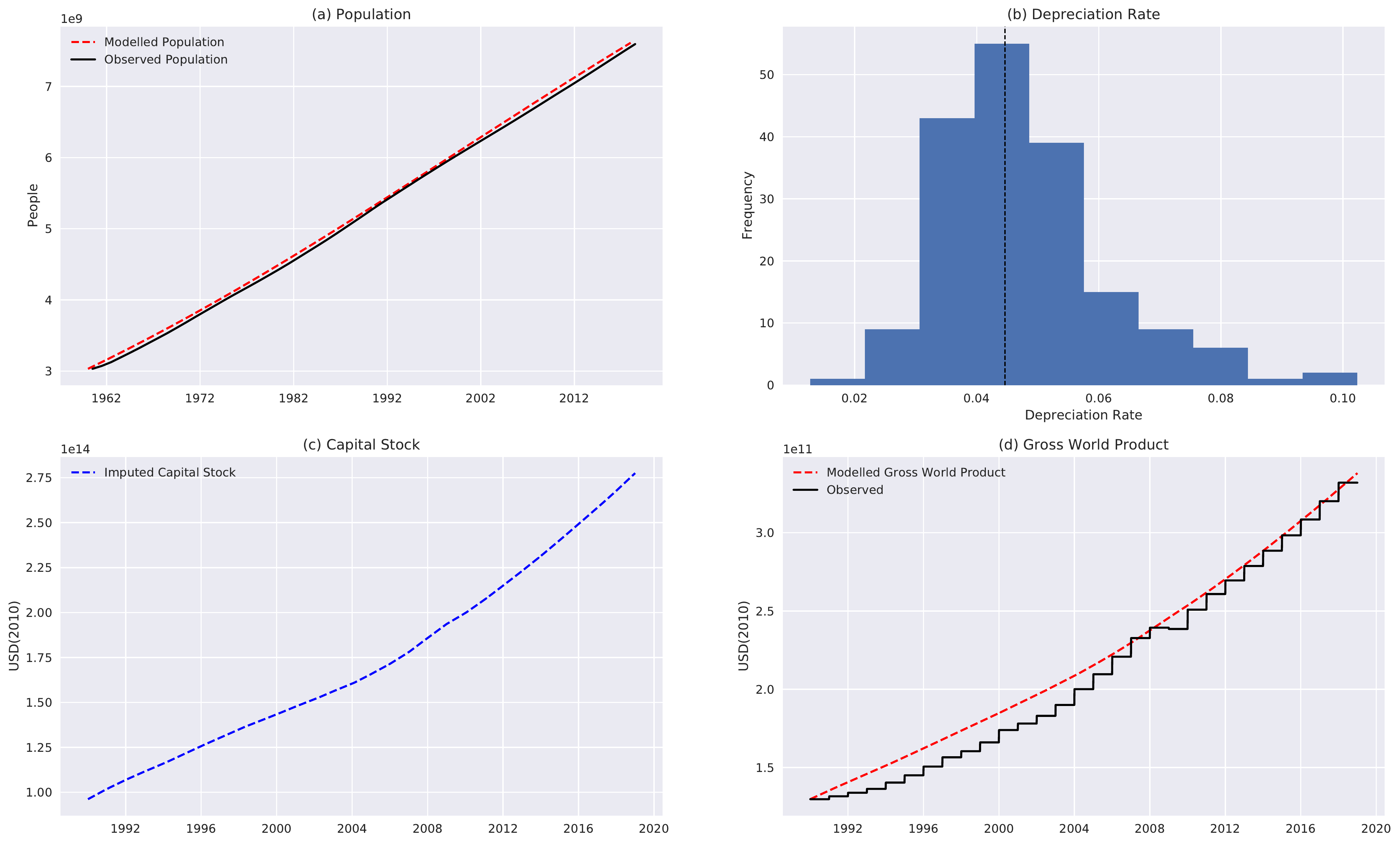}
    \caption{Calibration of the economic parameters. Panel (a): Global population and one-step-ahead predicted population from the fitted Gordon-Schaefer population growth model. Panel (b): National capital depreciation rates from the Penn World Tables 9.1 for 2017, with the dashed black line marking the median value $\delta=4.46\%$. Panel (c): Imputed values for daily global physical capital stock. Panel (d): Daily modelled gross world product, using a Cobb-Douglas production function with imputed daily capital stock, interpolated values for daily global population, output elasticity of capital set to 0.3, a growth rate for total factor productivity correspodning to 0.014 annually, and initial total factor productivity set to match actual production in 1990.}
    \label{fig:calibration}
\end{figure}

\paragraph{Population Growth} The parameters for the logistic population growth model, $a_1$ and $a_2$, were selected by first estimating a linear regression model on annual global population data from the World Bank between 1960 and 2018\footnote{Available at \url{https://data.worldbank.org/indicator/SP.POP.TOTL}, accessed on 2020-05-04}. The estimation results are shown in Table \ref{tab:pop}, and the regression coefficients $a_1^y$ and $a_2^y$ -- representing the parameters of an annual model -- were converted into their corresponding daily values by calculating:
\begin{gather*}
    a_1 = 1 + \frac{a^y_1-1}{365} \quad\quad a_2 = \frac{a^y_2}{365}.
\end{gather*}
The fitted values of the population model is shown in Panel (a) of Figure \ref{fig:calibration}, and appear to match the historical data for global population closely.

\begin{table} \centering 
  \caption{Estimated parameters for the Gordon-Schaefer population growth model. The model uses ordinary least squares, on annual data from 1960 to 2018.} 
  \label{tab:pop} 
\begin{tabular}{@{\extracolsep{5pt}}lc} 
\\[-1.8ex]\hline 
\hline \\[-1.8ex] 
 & \multicolumn{1}{c}{\textit{Dependent variable:}} \\ 
\cline{2-2} 
\\[-1.8ex] & World Population$_{t}$ \\ 
\hline \\[-1.8ex] 
 WorldPopulation$_{t-1}$ & 1.028$^{***}$ \\ 
  & (0.001) \\ 
  & \\ 
 World Population$_{t-1}^2$ & $-2.282 \times 10^{-12}$$^{***}$ \\ 
  & (0.000) \\ 
  & \\ 
\hline \\[-1.8ex] 
Observations & 58 \\ 
R$^{2}$ & 1.000 \\ 
Adjusted R$^{2}$ & 1.000 \\ 
Residual Std. Error & 4,813,932.000 (df = 56) \\ 
F Statistic & 36,877,908.000$^{***}$ (df = 2; 56) \\ 
\hline 
\hline \\[-1.8ex] 
\textit{Note:}  & \multicolumn{1}{r}{$^{*}$p$<$0.1; $^{**}$p$<$0.05; $^{***}$p$<$0.01} \\ 
\end{tabular} 
\end{table}

\paragraph{Capital Stock} We imputed the global physical capital stock by combining annual data on global gross physical capital formation, available for the period between from the World Bank\footnote{Available at \url{http://api.worldbank.org/v2/en/indicator/NE.GDI.TOTL.KD}, accessed on 2020-05-04.}, with an assumed physical capital depreciation rate of $\delta = 4.46\%$. This depreciation rate corresponds to the median value of the national depreciation rates for 2017 listed in the Penn World Tables 9.1\footnote{Available for download at \url{www.ggdc.net/pwt}, accessed on 2020-05-04.}, whose distribution is shown in Panel (b) of Figure \ref{fig:calibration}. The resulting estimated level of global physical capital stock from 1990 to 2019 is shown in Panel (c) of Figure \ref{fig:calibration}. 

\paragraph{Production} Following \cite{nordhaus_optimal_1992}, we set the output elasticity of capital to $\alpha = 0.3$. We then combine annual data of global output from the World Bank between 1990 and 2018\footnote{Available at \url{http://api.worldbank.org/v2/en/indicator/NY.GDP.MKTP.PP.KD}, accessed on 2020-05-04.} with the global population and the imputed global stock of physical capital, in order to estimate the total factor productivity, $A_t$, and its growth rate, $g$. This gives an annual total factor productivity growth rate of around $1.3\%$, with a corresponding daily growth rate of around $g=3.55 \times 10^{-5}$ over the period. The modelled global production is shown in Panel (d) of Figure \ref{fig:calibration}, and fits the observed data relatively well, although the modelled production level slightly overestimates global production at some points.

\paragraph{Utility} We select an annual utility discount rate that corresponds to an annual rate of $\rho=8\%$. This discount rate allows the simulated investment from the model to match the observed gross physical capital formation in the period between 1990 and 2010, as shown in Panel (d) of Figure \ref{fig:backtest}. Although this discount rate appears somewhat high, it is not unreasonable if we take into consideration that the model represents the global economy, and that large parts of the global population consists of low-income households with high discount rates.

\begin{figure}
    \centering
    \includegraphics[width=0.5\textwidth]{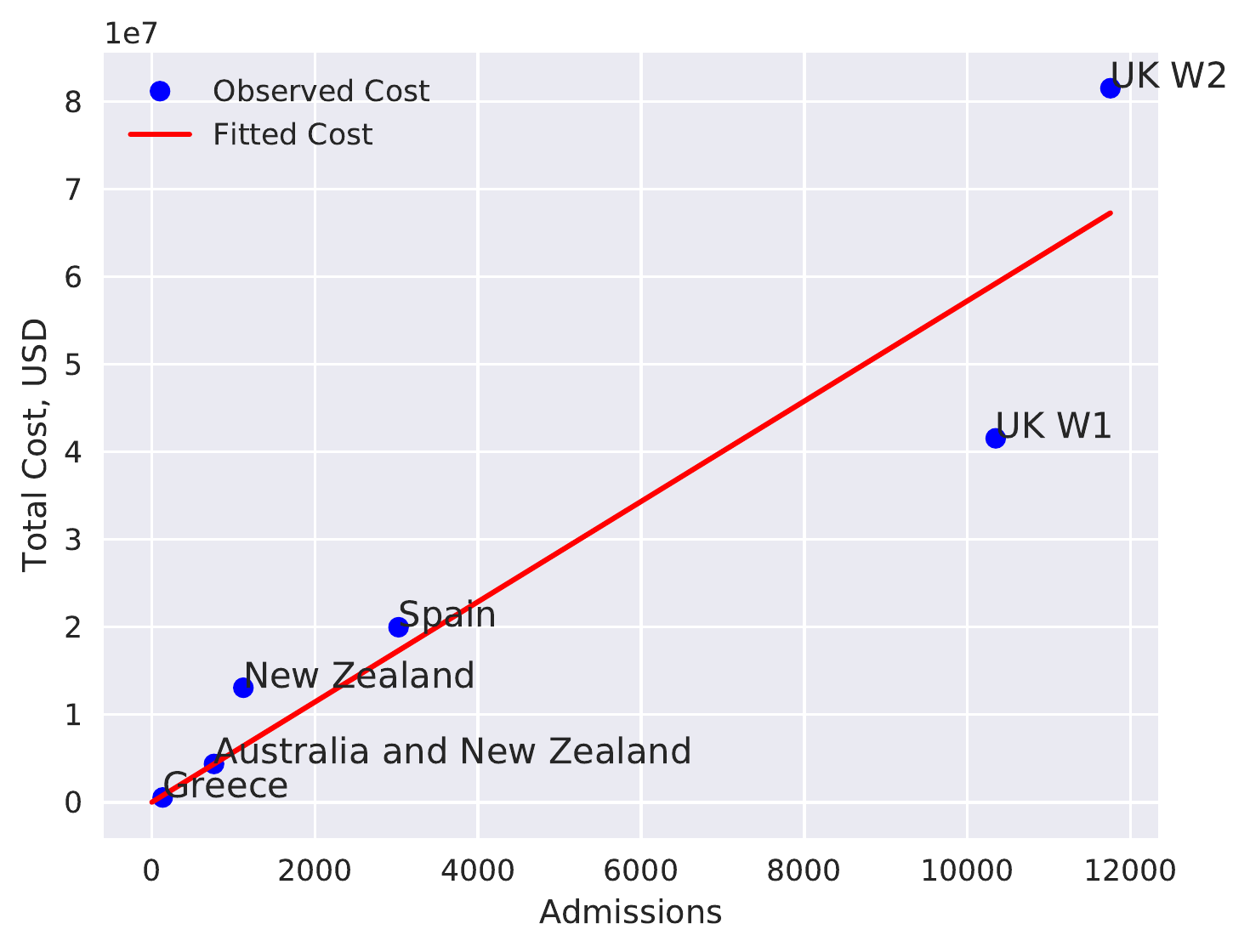}
    \caption{Direct costs of the H1N1 pandemic, based on data compiled by \cite{lau_excess_2019}.}
    \label{fig:costs}
\end{figure}

\paragraph{Excess Direct Pandemic Costs} A pandemic directly causes additional costs to society, which is captured by the function $H(\cdot)$ in our mathematical model. To model this cost, we look to the literature which has estimated the excess hospital admission costs for a recent similar pandemic, the H1N1 pandemic in 2009: for Spain \citep{galante_health_2012}, Greece \citep{zarogoulidis_health_2012}, Australia and New Zealand \citep{higgins_critical_2011}, New Zealand \citep{wilson_national_2012}, and the United Kingdom \citep{lau_excess_2019}. Figure \ref{fig:costs} shows the direct hospitalisation costs attributed to the H1N1 pandemic in various countries, along with the number of hospital admissions. Based on these previous cost estimates for the H1N1 pandemic, we use a flat cost of $u = 5,722$ USD per hospital admission (see Table \ref{tab:direct_costs}), corresponding to the solid red line in Figure \ref{fig:costs}. Although we assume a flat cost per admission, it may be more reasonable in other contexts -- for instance when applying the model to specific regions -- to consider a cost function with an increasing marginal cost: as hospital capacity becomes constrained in the short-run during a surge of admissions, one could expect the unit cost to increase. However, in the context of our global model, we do not distinguish between the regions in which the cases occur, and therefore cannot accurately capture such a saturation effect. Therefore, we choose a direct cost function that is simply linear in the number of hospital admissions.

We assume that $h=14.7\%$ of the confirmed infected cases will be admitted to hospital\footnote{This hospitalisation rate corresponds to the median of USA state level hospitalisation rates reported in the daily COVID-19 reports from the Center for System Science and Engineering at John Hopkins University, on the 10\textsuperscript{th} of May 2020, available at \url{https://github.com/CSSEGISandData/COVID-19}.}, and our direct cost function is given by:
\begin{gather*}
    H_t = u h b S_t I_t.
\end{gather*}

\begin{table} \centering 
  \caption{Esimated direct cost function.} 
  \label{tab:direct_costs} 
\begin{tabular}{@{\extracolsep{5pt}}lc} 
\\[-1.8ex]\hline 
\hline \\[-1.8ex] 
 & \multicolumn{1}{c}{\textit{Dependent variable:}} \\ 
\cline{2-2} 
\\[-1.8ex] & Total Costs (USD) \\ 
\hline \\[-1.8ex] 
 Admissions & 5,722.078$^{***}$ \\ 
  & (664.874) \\ 
  & \\ 
\hline \\[-1.8ex] 
Observations & 6 \\ 
R$^{2}$ & 0.937 \\ 
Adjusted R$^{2}$ & 0.924 \\ 
Residual Std. Error & 10,643,570.000 (df = 5) \\ 
F Statistic & 74.068$^{***}$ (df = 1; 5) \\ 
\hline 
\hline \\[-1.8ex] 
\textit{Note:}  & \multicolumn{1}{r}{$^{*}$p$<$0.1; $^{**}$p$<$0.05; $^{***}$p$<$0.01} \\ 
\end{tabular} 
\end{table} 

\begin{figure}
    \centering
    \includegraphics[width=\textwidth]{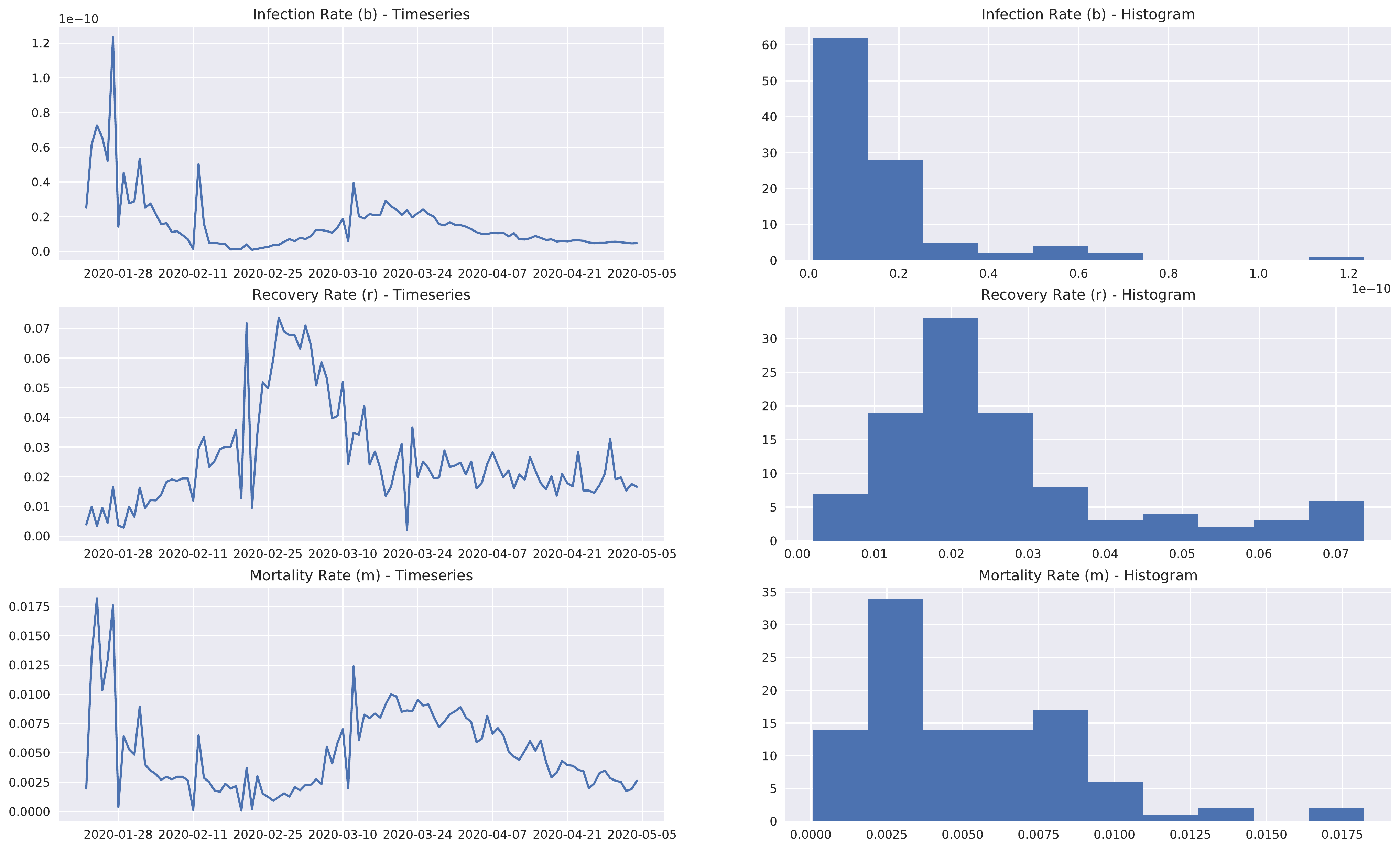}
    \caption{Calibration of the parameters for the SIR model. Panel (a) and (b): the development and distribution of the SIR infection rate, $b$. Panel (c) and (d): the development and distribution of the SIR recovery rate, $r$. Panel (e) and (f): the development and distribution of the SIR mortality rate, $m$.}
    \label{fig:epidemicparameters}
\end{figure}

\paragraph{Infection, Recovery and Mortality Rates} To estimate the mortality and recovery rates, $r$ and $m$, we solve the transition equations for the number of recovered $R_t$ (equation \eqref{eq:recovere}) and the number of deceased $D_t$ (equation \eqref{eq:deceased}) for their respective parameters:
\begin{gather*}
    r = \frac{R_{t+1}-R_t}{I_t}, \quad m = \frac{D_{t+1} - D_t}{I_t}.
\end{gather*}
Using daily data for the number of confirmed, recovered and deceased cases, made available by John Hopkins University\footnote{Available at \url{https://github.com/CSSEGISandData/COVID-19}, accessed on 2020-05-06.}, we can calculate the recovery and mortality rates for each day, as shown in the bottom two rows of Figure \ref{fig:epidemicparameters}.

To estimate the infection rate, $b$, we solve equation \eqref{eq:infected} for the parameter $b$:
\begin{gather*}
    b = \frac{I_{t+1} - (1-r-m)I_t}{S_t I_t}.
\end{gather*}
Taking into account that $I_t$ in the model refers to the number of active cases, whereas the data reports the accumulated number of cases, and using the population growth model to help estimate the number of susceptible individuals, we calculate the daily infection rates, shown in the top row of Figure \ref{fig:epidemicparameters}. As the infection rate $b$ varies over time, we choose a relatively high infection rate to represent the infection rate in the absence of policy intervention, $b_0 = 2.041\times 10
^{-11}$, which equals the upper quartile (75\%) of the observed infection rates. As we simulate different intervention policies, this base infection rate $b_0$ will be modified.

\begin{figure}
    \centering
    \includegraphics[width=0.5\textwidth]{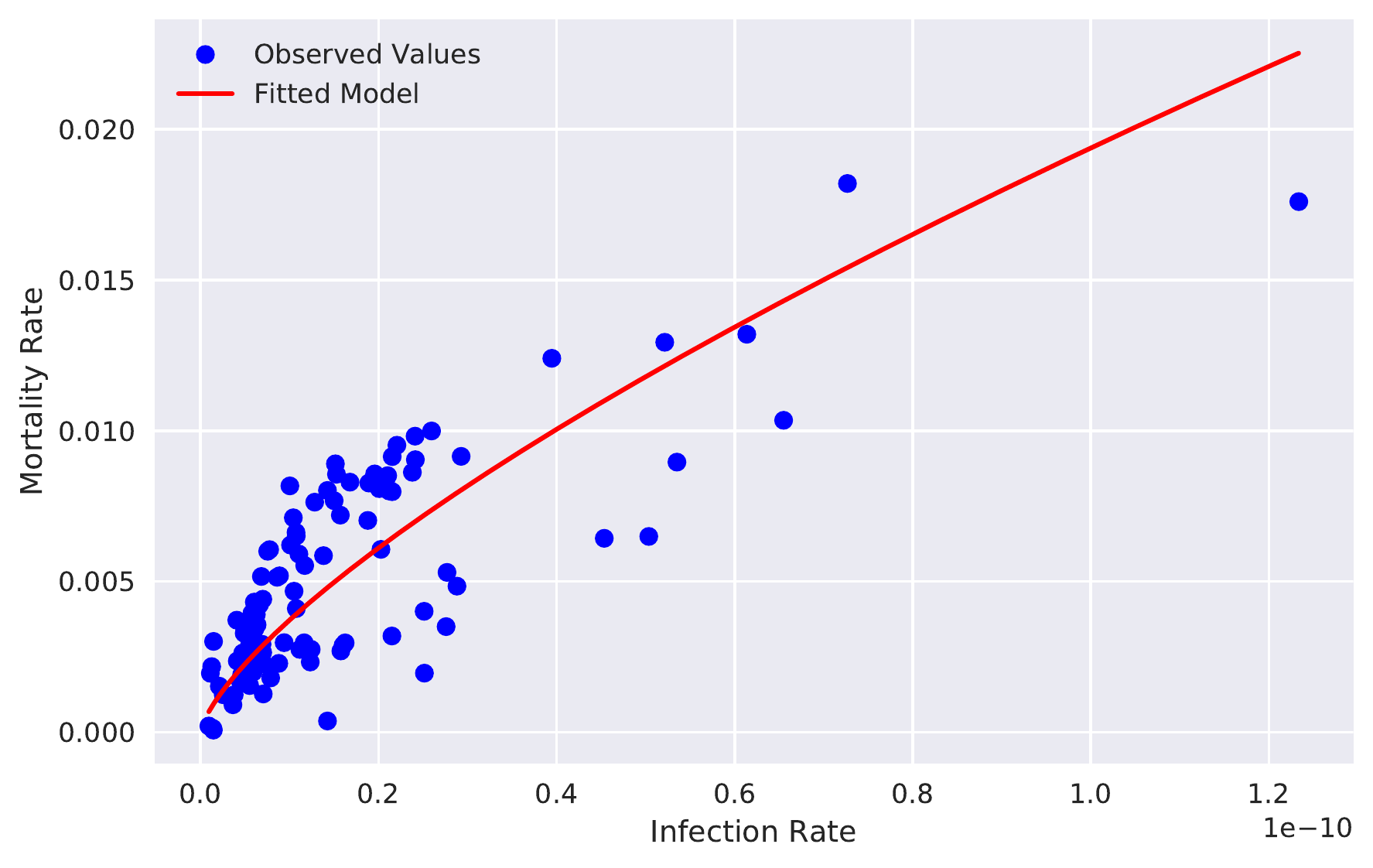}
    \caption{Global mortality rate and infection rate.}
    \label{fig:mortalityrate}
\end{figure}

We notice from Figure \ref{fig:epidemicparameters} that the mortality rate appears to rise and fall together with the infection rate\footnote{\cite{alvarez_simple_2020} have also made this observation, and included the effect in their model.}. This might reflect that lack of capacity in a health system increases mortality rate, and this is therefore a feature that we would like to capture. We therefore estimate $m$ as a function of $b$:
\begin{gather*}
    m = k_1 b^{k_2},
\end{gather*}
in which $k_1$ and $k_2$ are constants. Table \ref{tab:mortality} shows the regression for estimating the parameters $k_1$ and $k_2$, and the fitted function is shown in Figure \ref{fig:mortalityrate}, together with the observed values of daily infection and mortality rates.

\begin{table} \centering 
  \caption{Mortality rate model.}
  \label{tab:mortality} 
\begin{tabular}{@{\extracolsep{5pt}}lc} 
\\[-1.8ex]\hline 
\hline \\[-1.8ex] 
 & \multicolumn{1}{c}{\textit{Dependent variable:}} \\ 
\cline{2-2} 
\\[-1.8ex] & ln(Mortality Rate) \\ 
\hline \\[-1.8ex] 
 ln(Infection Rate) & 0.717$^{***}$ \\ 
  & (0.065) \\ 
  & \\ 
 Constant & 12.561$^{***}$ \\ 
  & (1.642) \\ 
  & \\ 
\hline \\[-1.8ex] 
Observations & 104 \\ 
R$^{2}$ & 0.545 \\ 
Adjusted R$^{2}$ & 0.540 \\ 
Residual Std. Error & 0.619 (df = 102) \\ 
F Statistic & 122.131$^{***}$ (df = 1; 102) \\ 
\hline 
\hline \\[-1.8ex] 
\textit{Note:}  & \multicolumn{1}{r}{$^{*}$p$<$0.1; $^{**}$p$<$0.05; $^{***}$p$<$0.01} \\ 
\end{tabular} 
\end{table} 

For the recovery rate in the simulations, we select the median of the daily recovery rates calculated from the data, $r=0.02099$, which implies that slightly over 2\% of infected individuals recover from one day to the next. The infection rate, $b$, however, will be determined individually for each scenario, and will reflect the pandemic policy simulated in each of the scenarios. The mortality rate $m$ will be determined by the infection rate $b$, according to the relationship between them we estimated earlier.

\paragraph{The Production-Infection Trade-Off} Our model assumes that pandemic policy mainly impacts the spread of the pandemic through manipulating the infection rate $b$, and mainly infects economic growth by affecting production of economic output, $Y_t$. Our model contains a single parameter, $p$, to represent pandemic policy, which directly represents the shortfall in global production. In order to analyse the trade-off between production and the infection rate, however, we must establish how the infection rate, $b$, is impacted by the policy parameter $p$ -- that is, we must quantify how the infection rate responds to foregone production.

We expect the relationship between the infection rate, $b$, and the GDP shortfall, $p$, to exhibit two specific characteristics. Firstly, we expect a reduction in the infection rate as the GDP shortfall increases, because we assume pandemic policies are designed to reduce the infection rate, which result in a shortfall in the economic production as a side-effect. Secondly, we expect the reduction in the infection rate to be greater at first, because we expect measures to be enacted in order from more to less effective, and from less to more disruptive. That is, the infection rate reductions exhibit a form of decreasing returns in the GDP shortfall. We suggest that the percentage reduction in the infection rate, $\Delta b(\%)$, responds to the percentage reduction in GDP, $\Delta \text{GDP}(\%)$, as follows:
\begin{gather*}
    \Delta b(\%) = q_1 \Delta \text{GDP}(\%)^{q_2},
\end{gather*}
in which $q_1$ and $q_2$ are constants, and $q_2 \in (0,1)$.

The shortfall in production associated with each infection rate is, however, not straight-forward to estimate. Firstly, GDP data is published several months late, which means that a whole pandemic might have long since passed by the time the GDP data is published. Secondly, GDP data are generally aggregated into months, quarters or years, which would make it difficult to associate with a particular infection rate, which can vary greatly over these timeframes. Therefore, we infer the daily reduction in economic production from the shortfall in electricity consumption: electricity consumption data for many countries is available near real-time -- often at sub-hourly resolution -- and electricity consumption is known to correlate well with economic activity (see, for instance, \cite{trotter_climate_2016} and \cite{rodriguez_climate_2019}).

Data on electricity demand (load) is available for many European countries from the ENTSO-E Transparency Platform\footnote{Available at \url{https://transparency.entsoe.eu/}, accessed on 2020-05-13.}, and we utilise data from Europe's five biggest economies, which have all been significantly impacted by COVID-19: France, Germany, Italy, Spain, and the United Kingdom. To measure the daily \textit{shortfall} in electricity consumption, however, we must compare the observed electricity consumption to what it would have been under normal conditions. Therefore, we first need to create a counterfactual representing the electricity consumption under normal conditions. We use the automated forecasting procedure by \cite{taylor_forecasting_2017} to calibrate models on the daily national electricity consumption (load) data from 2015 until March 1, 2020. This period does not include the main impacts of the pandemic, such that forecasts from the models for the period from March 1, 2020 to May 10, 2020 can act as counterfactuals -- how electricity consumption would have been expected to develop under normal conditions. These counterfactuals may then be compared to the observed values in the same period, and allows us to calculate the daily electricity consumption shortfall in terms of percentages.

\begin{figure}
    \centering
    \includegraphics[width=\textwidth]{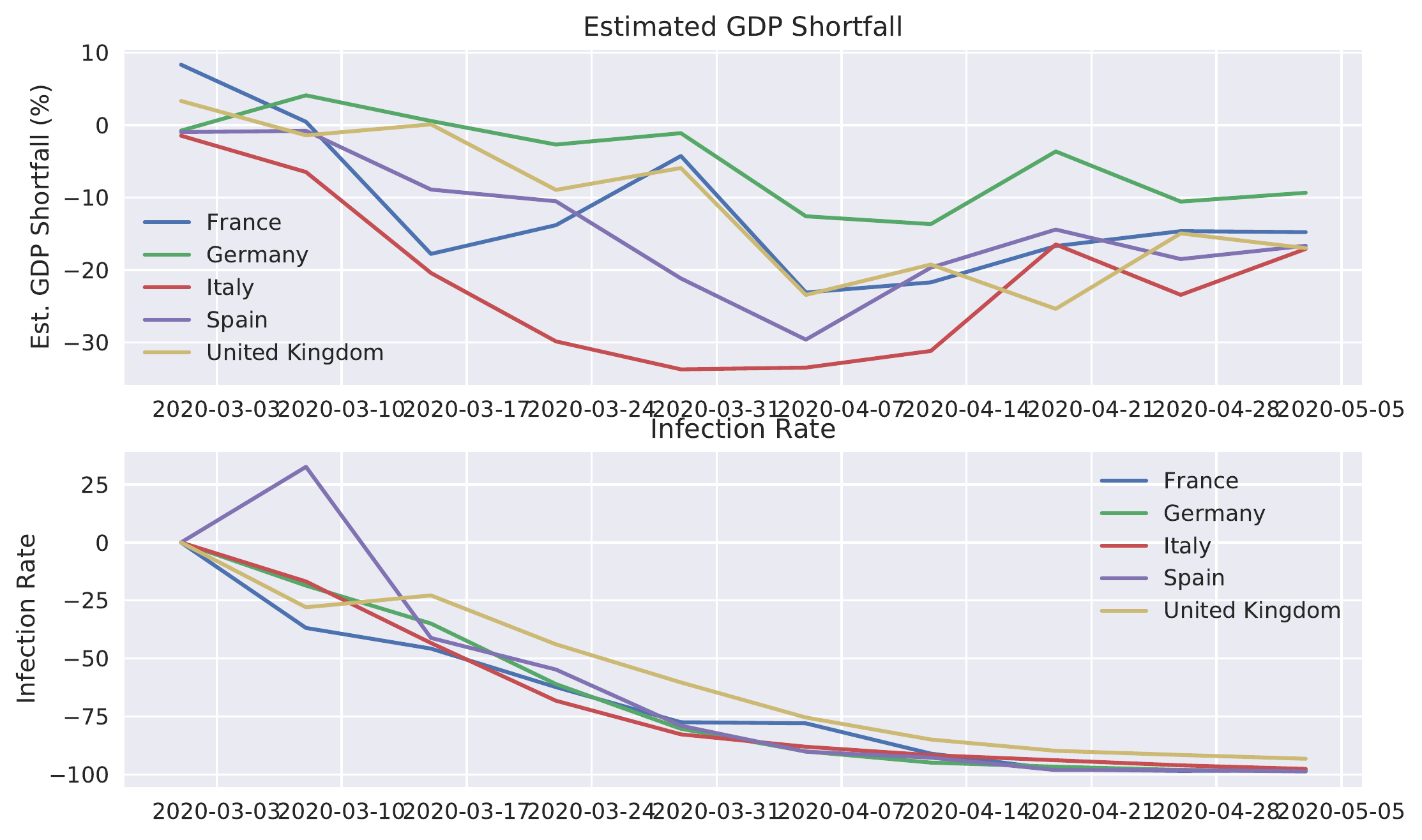}
    \caption{Weekly estimated GDP shortfall for France, Germany, Italy, Spain, and the United Kingdom. The GDP shortfall is based on the difference between actual and expected electricity consumption. The average weekly infection rate is based on the number of confirmed cases.}
    \label{fig:prodinfecttime}
\end{figure}

\begin{figure}
    \centering
    \includegraphics[width=0.5\textwidth]{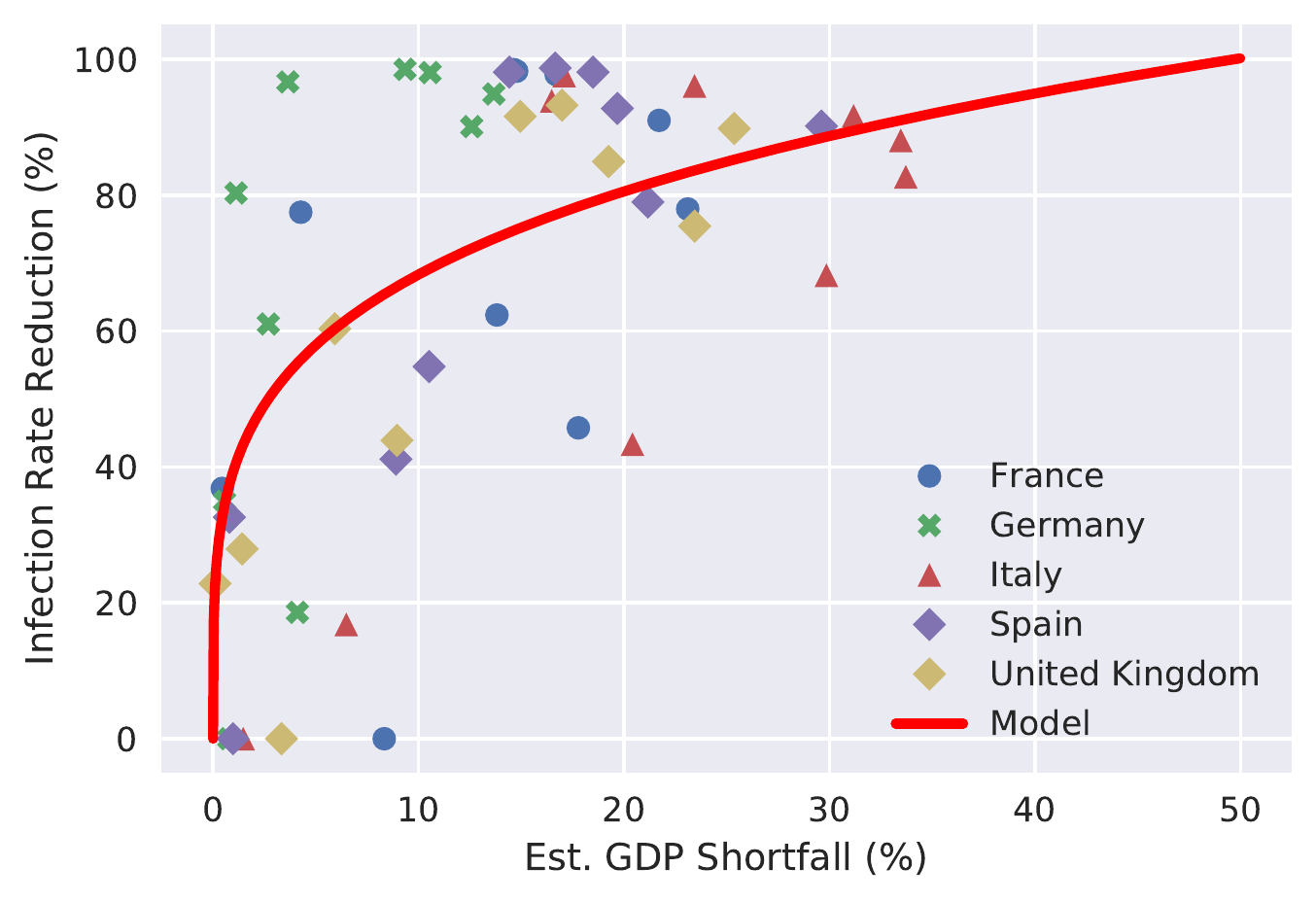}
    \caption{Weekly estimated GDP shortfalls for France, Germany, Italy, Spain, and the United Kingdom, based on the difference between actual and expected electricity consumption. The average weekly infection rate is based on the number of confirmed cases. The red line represents the model estimated on the data.}
    \label{fig:prodinfectscatter}
\end{figure}

The relationship between electricity consumption and production has been the subject of many studies (often as variants of ``income elasticity of electricity consumption''), and, synthesising the studies into a useful heuristic, we assume that a 1\% decrease in electricity consumption is associated with a 1.5\% reduction in GDP. The estimated GDP shortfall for France, Germany, Italy, Spain and the United Kingdom are illustrated in Panel (a) of Figure \ref{fig:prodinfecttime}, which shows a clear increase in the GDP shortfall throughout March 2020, and a stable shortfall of around 10\%-20\% throughout April and the start of May 2020, appearing to correspond closely to the lockdown periods of these countries. Panel (b) of Figure \ref{fig:prodinfecttime} shows the reduction in the infection rate for the five countries over the same time period, with the base period for the infection rate considered to be the first seven days in March, and it is clear that the infection rate has decreased as the GDP shortfall has dropped, which is consistent with our expectations. The scatter plot of the estimated GDP shortfall and the reduction in infection rates, shown in Figure \ref{fig:prodinfectscatter}, shows a clear relationship between infection rate reductions and GDP shortfall. The red line in Figure \ref{fig:prodinfectscatter} shows the model, with constants estimated as in Table \ref{tab:infxgdp}. The model fits the observations well, and the estimated values for the constants, $q_1$ and $q_2$, conform to our expectations.

\begin{table} \centering 
  \caption{Model of reductions in the infection rate as a function of reductions in economic production.}
  \label{tab:infxgdp} 
\begin{tabular}{@{\extracolsep{5pt}}lc} 
\\[-1.8ex]\hline 
\hline \\[-1.8ex] 
 & \multicolumn{1}{c}{\textit{Dependent variable:}} \\ 
\cline{2-2} 
\\[-1.8ex] & $\ln(\Delta b(\%))$ \\ 
\hline \\[-1.8ex] 
 $\ln(\Delta\text{GDP} (\%))$ & 0.238$^{***}$ \\ 
  & (0.045) \\ 
  & \\ 
 Constant & 3.677$^{***}$ \\ 
  & (0.114) \\ 
  & \\ 
\hline \\[-1.8ex] 
Observations & 45 \\ 
R$^{2}$ & 0.398 \\ 
Adjusted R$^{2}$ & 0.384 \\ 
Residual Std. Error & 0.383 (df = 43) \\ 
F Statistic & 28.435$^{***}$ (df = 1; 43) \\ 
\hline 
\hline \\[-1.8ex] 
\textit{Note:}  & \multicolumn{1}{r}{$^{*}$p$<$0.1; $^{**}$p$<$0.05; $^{***}$p$<$0.01} \\ 
\end{tabular} 
\end{table}

\paragraph{} Table \ref{tab:parameters} summarises the chosen values for the parameters in the model. Having defined parameter values such that the model represents the global economy under the impact of COVID-19, we can now define scenarios that can be simulated numerically, and provide insight into the impact of policy on both economic growth and the spread of the pandemic.

\begin{table}
    \centering
    \caption{Parameter values.}
    \resizebox{\columnwidth}{!}{%
    \begin{tabular}{llrl}
		\toprule
         \textbf{Parameter} & \textbf{Description} & \textbf{Value} \\
		\midrule
		 $a_1$, $a_2$ & Logistic population growth (annual) & 1.028, -2.282$\times$10$^{-12}$ \\
		 $\delta$ & Capital depreciation rate (annual) & 4.46\% \\
         $\alpha$ & Output elasticity of capital & 0.3 \\
         $g$ & Growth rate of total factor productivity (annual) & 1.3\% \\
         $\rho$ & Utility discount rate (annual) & 8\% \\
         $u$ & Cost per hospital admission & 5,722 USD \\
         $h$ & Hospital admissions per confirmed case & 14.7\% \\
         $r$ & Daily recovery rate per active infection & 2.1\% \\
         $b_0$ & Base infection rate (no intervention) & $2.041 \times 10^{-11}$ \\
         $k_1, k_2$ & Mortality rate parameters, $m = k_1 b^{k_2}$ & 12.561, 0.717 \\
         $q_1, q_2$ & Infection rate parameters, $\Delta b(\%) = q_1 \Delta\text{GDP}(\%)^{q_2}$ & 3.677, 0.238 \\
		\bottomrule
    \end{tabular}
    }
    \label{tab:parameters}
\end{table}

\subsection{Policy Experiments}
To have a basis for comparison, we first simulate two baseline scenarios: the \textit{No Pandemic} scenario, in which no pandemic occurs, and the \textit{No Intervention} scenario, in which the pandemic occurs with no direct intervention ($p=0$). Comparing the remaining scenarios to the first baseline scenario (\textit{No Pandemic}) will help us understand the impact of the pandemic. In addition, the baseline scenario will provide the initial conditions for population and capital stock at the start of the pandemic, as observations are not yet available. Comparing the remaining scenarios to the second baseline scenario (\textit{No Intervention}) will help us understand the impact of the simulated policy intervention. The initial values for these simulations are shown in Table \ref{tab:baselines}.

Having established the baseline scenarios, we run a series of simulations to investigate three fundamental aspects of the policy intervention. First, we alter the timing of the start of the intervention, to explore the advantages and disadvantages of starting the intervention early or late. Second, we alter the intensity of the intervention, to investigate the differences in the impacts between light and severe interventions. And, finally, we alter the duration of the intervention. That is, by running numerical experiments that vary the policy interventions in commencement, intensity and duration, we answer three fundamental policy questions: ``When?'', ``How much?'', and ``For how long?''. When taken together, these experiments will provide insight into the economic and health impacts of varying policies along these three dimensions, and highlight the trade-offs that policymakers must consider:
\begin{description}
    \item[When to intervene?] Holding the intervention intensity and duration fixed at 10\% and 26 weeks, we run simulations altering the start of the policy intervention between April 09, 2020, and June 02, 2020.
    \item[How much?] Holding the starting date of the intervention fixed at March 12, 2020 -- the date when the WHO declared COVID-19 to be a pandemic -- and the duration fixed at 26 weeks, we alter the intensity of the intervention from 5\% to 25\%, in steps of 10 percentage points. 
    \item[For how long?] Keeping the starting date of the intervention fixed at March 12, 2020, and the intervention intensity fixed at 10\%, we alter the duration of the intervention between 4 weeks and 76 weeks.
\end{description}
The initial values used in all these simulations are the same as in the \textit{No Intervention} scenario, specified in Table \ref{tab:baselines}. Taken together, these three sequences of simulations will provide important and actionable insights into the impacts of policy intervention on both economic growth and on the spread of the pandemic that will help policymakers understand the relevant trade-offs.

\begin{table}
    \centering
    \caption{Parameters used for the baseline scenarios.}
    \resizebox{\columnwidth}{!}{%
    \begin{tabular}{llrrrrrrrrrr}
		\toprule
		\midrule
        \textbf{Scenario} & \textbf{Start date} & $N_0$ & $I_0$ & $R_0$ & $D_0$ & $b_0$ & $A_0$ & $K_0$ \\
		\midrule
        \textbf{No Pandemic} & 2019-01-01 & 7.634$\times 10^{9}$ & 0 & 0 & 0& 0 & 1.880 & 2.775$\times 10^{14}$ \\
        \textbf{No Intervention} & 2020-01-22 & 7.718$\times 10^9$ & 510 & 28 & 17 & 2.041$\times 10^{-11}$ & 1.906 & 2.827$\times 10^{14}$ \\
		\bottomrule
    \end{tabular}}
    \label{tab:baselines}
\end{table}
    
\section{Results and Discussion}
\subsection{Backtest 1990-2010}
\begin{figure}
    \centering
    \includegraphics[width=\textwidth]{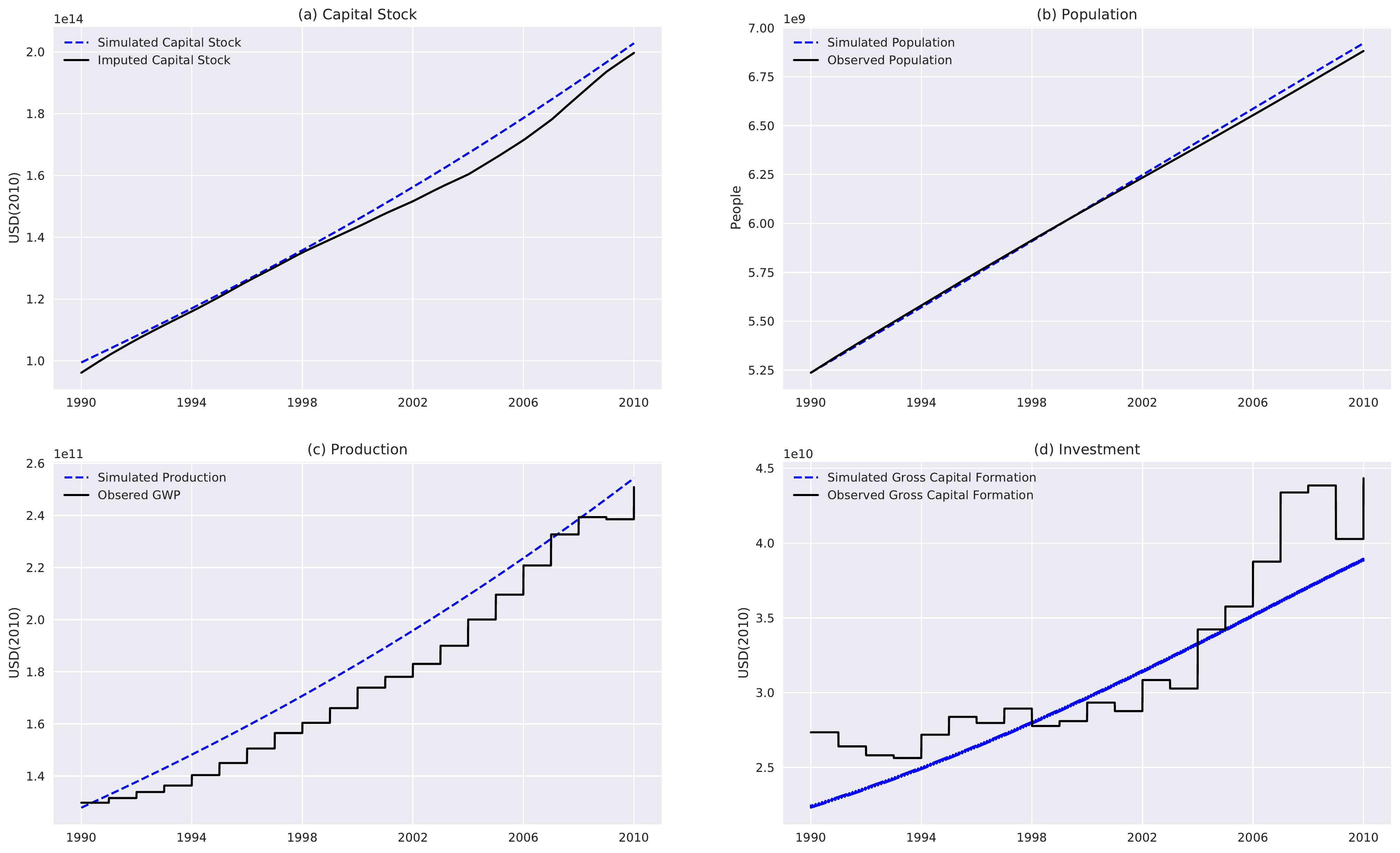}
    \caption{Model backtest results. Panel (a): The simulated daily development of the global physical capital stock and the daily imputed global physical capital stock. Panel (b): Global daily simulated and observed population. Panel (c): Simulated and observed daily gross world production. Panel (d): Simulated and observed daily global gross physical capital formation.}
    \label{fig:backtest}
\end{figure}

Before presenting the main simulation results, we first present the results of a backtest. This is shown in Figure \ref{fig:backtest}, and shows that the model captures the main features of the observed historical data. Although the backtest in this case is not an out-of-sample test, due to lack of data, the backtest provides strong support for the economic components of the model.

\subsection{Baseline Scenarios}
\begin{figure}
    \centering
    \includegraphics[width=\textwidth]{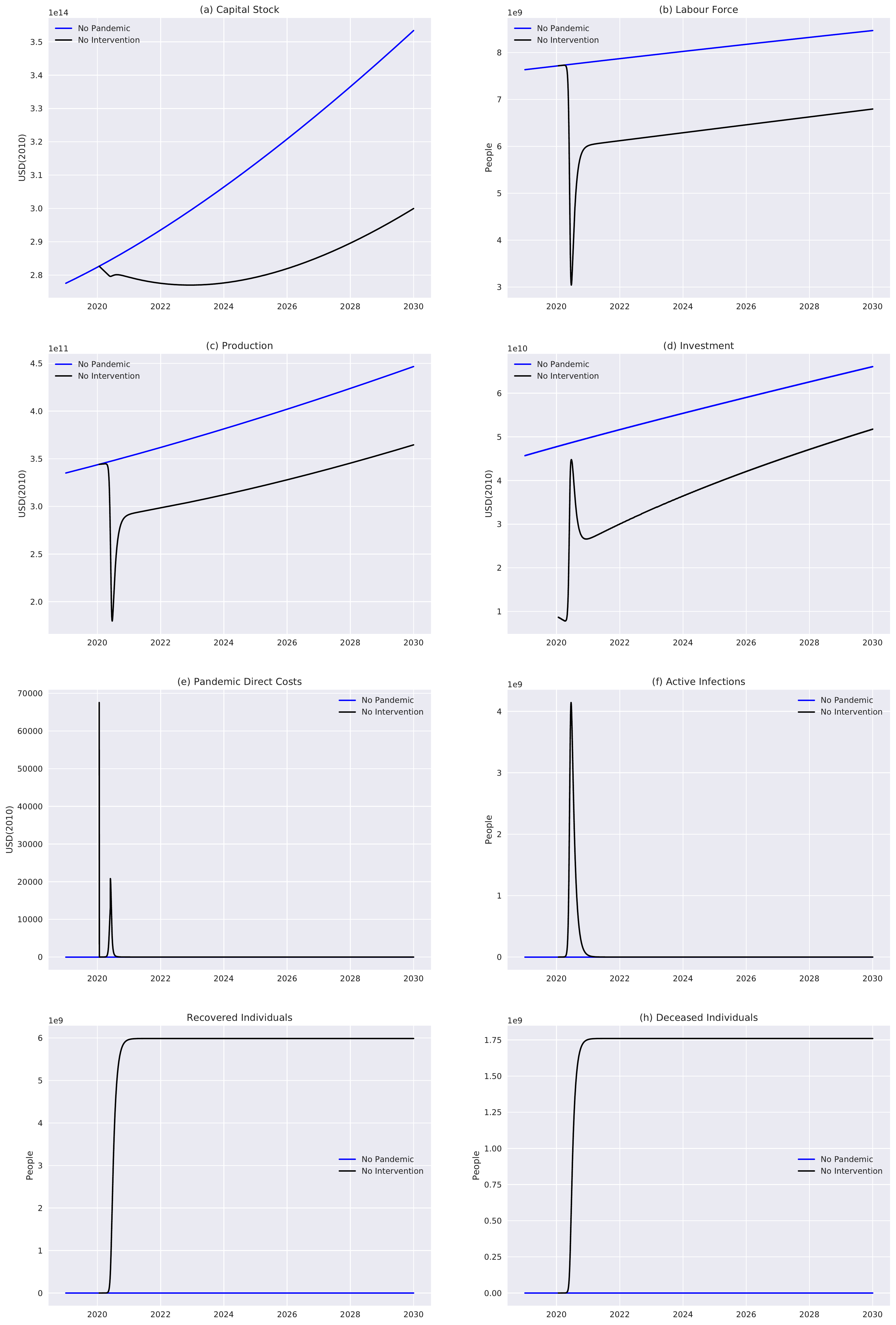}
    \caption{Simulation results for the baseline scenarios.}
    \label{fig:baselines}
\end{figure}

The results of simulating the baseline scenarios -- \textit{No Pandemic} and \textit{No Intervention} -- are shown in Figure \ref{fig:baselines}. As expected, the \textit{No Pandemic} scenario is characterised by steady economic growth, and no infected or deceased individuals. The \textit{No Intervention} scenario, however, shows a large and abrupt drop of around 45\% in production during the first half of 2020, as the pandemic spreads through the population. The number of active infections peaks in mid-June, 2020. As the pandemic subsides, a large proportion of the labour force never returns as the mortalities reach 1.75 billion people, and production recovers only to 85\% of its pre-pandemic value before 2021. Although growth in economic production resumes after the pandemic, production remains 20\%-25\% below the production in the \textit{No Pandemic} scenario until the end of the simulation in 2030. In summary, the \textit{No Intervention} scenario shows substantial loss of human life, as well as a lasting and significant negative impact on production, and we expect that shrewd policy intervention could partially mitigate these impacts.

In the following, we run model simulations to gain insight into when to start the policy intervention, to what degree to intervene, and for how long the intervention should last.

\subsection{When to Intervene?}
\begin{figure}
    \centering
    \includegraphics[width=\textwidth]{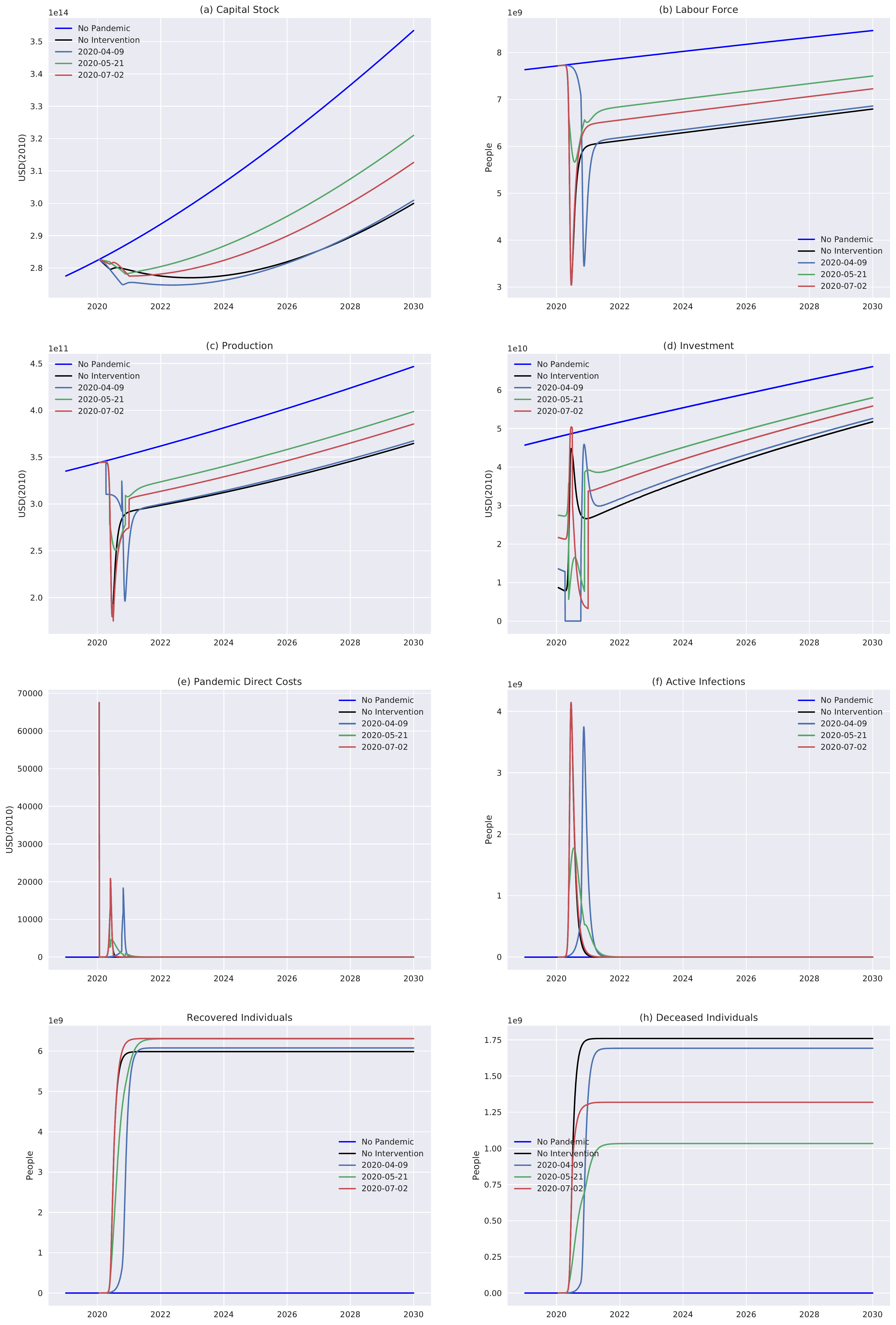}
    \caption{Simulation results when varying the starting date of the policy intervention.}
    \label{fig:startdates}
\end{figure}

We first run a series of simulations to examine the question of when a possible policy intervention should start. In this series of simulations, the intervention intensity is held fixed at 10\% (that is, the intervention causes a 10\% decline in production), and the duration of the intervention is held fixed at 26 weeks. Multiple simulations are run with differing starting dates for the policy intervention. This series of simulations is shown in Figure \ref{fig:startdates}, with three possible starting dates for the policy intervention: April 9, May 21, and July 2.

Examining Panel (f) of Figure \ref{fig:startdates}, we note that intervening on July 2 allows the pandemic to spread almost identically to the \textit{No Intervention} scenario -- that is, July 2 is too late for effective intervention because the peak in active infections has passed, most of the damage is already done, and the pandemic is decelerating by itself. However, by intervening on July 2 when the number of active infections is near its highest, many mortalities are avoided, and the human and economic damage is somewhat lower than in the \textit{No Intervention} scenario. Further, we note that intervening on April 2 does not appear to significantly alter the course of the pandemic or mitigate its effects -- intervening so early in the pandemic only serves to delay the main wave of infections.

The intervention starting on May 21 -- about one month before the peak of the \textit{No Intervention} scenario -- appears to be the most effective of our simulations, both considering the economic impacts and the final mortality rate. May 21 appears to be just before the inflection point of the \textit{No Intervention} scenario, and the number of infections is growing at its highest rate. Between the three simulated scenarios, this is by far the preferred option.

It seems that timing the policy intervention is of great importance to mitigate both the human and the economic impacts. Although we do not believe that the exact dates hold for the COVID-19 pandemic in particular, these simulations lead us to interesting insights: policy intervention appears to be most effective when the number of active infections is approaching its inflection point, and is growing at its highest rate. An intervention that is too early will only serve to delay the critical phase of the pandemic, and an intervention after the peak has occurred will obviously do nothing to lower the peak. Although it may be difficult to know beforehand when a pandemic will enter its critical phase, the timing of the policy intervention is of paramount importance, and our results suggest that authorities should implement emergency policies only when the disease is sufficiently spread. However, we expect that the exact specification of \textit{sufficiently spread} to vary significantly from place to place, depending on local conditions.

This finding contradicts the claims of \cite{guan_global_2020}, who argue that interventions should be ``earlier, stricter and shorter'': instead, our results show that starting the intervention before the disease is sufficiently spread will either simply delay the critical phase of the pandemic (if the intervention indeed is kept ``short''), or prolong the intervention (if the intervention is extended). The study by \cite{eichenbaum_macroeconomics_2020} focuses on ``starting too late'' and ``ending too early'', yet our results suggest that policymakers also need to avoid starting \emph{too} early.

\subsection{How Much Intervention?}
\begin{figure}
    \centering
    \includegraphics[width=\textwidth]{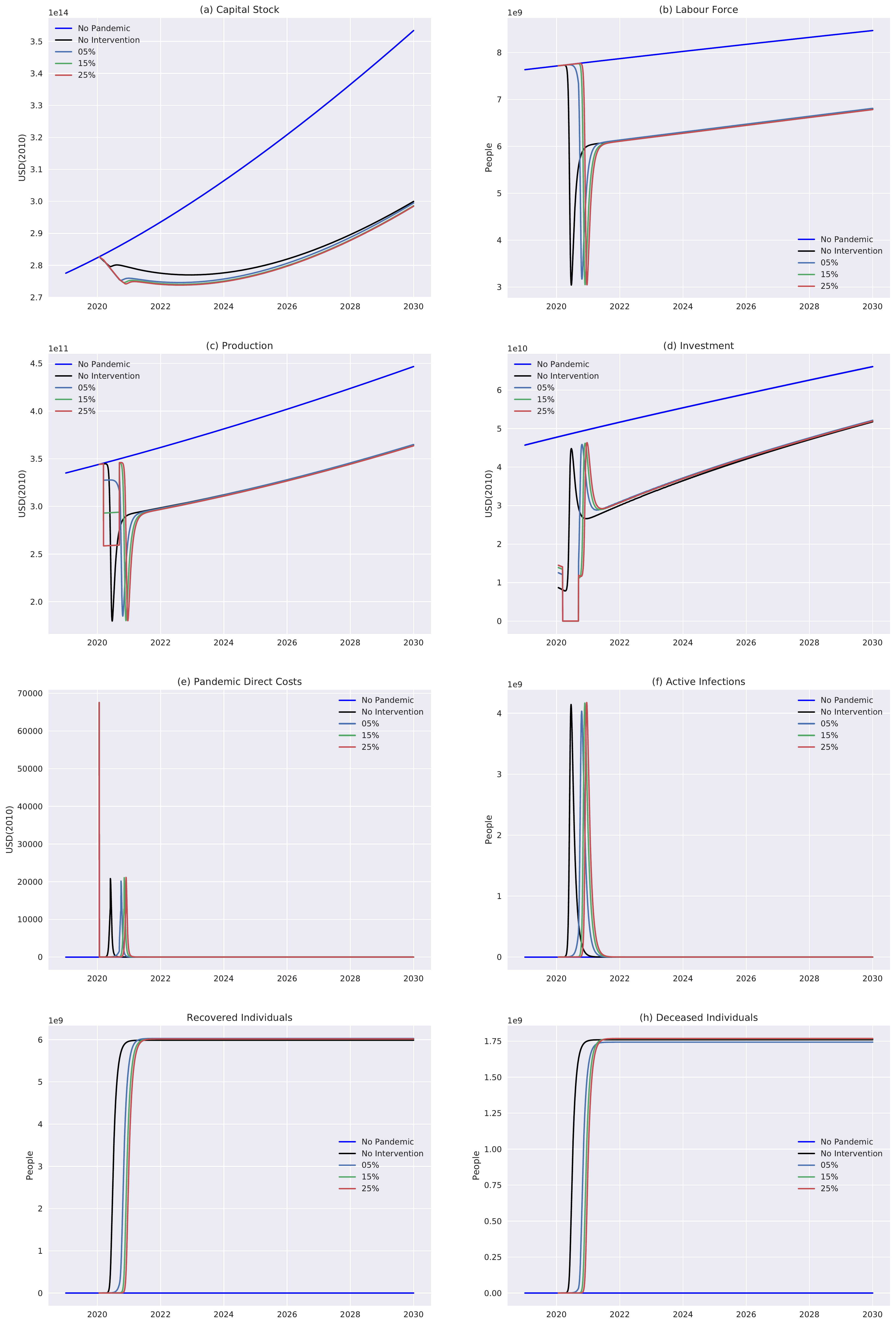}
    \caption{Simulation results when varying the intensity of the policy intervention.}
    \label{fig:degrees}
\end{figure}

In this series of simulations, we keep the starting date of the policy intervention fixed at March 12 and the duration of the policy intervention fixed at 26 weeks, whilst varying the intervention inensity. We simulate policies that reduce production by 5\%, 15\%, and 25\%, and the simulation results are shown in Figure \ref{fig:degrees}.

The three simulations with different intervention intensity, shown in Figure \ref{fig:degrees}, suggest that varying the intensity of the intervention mainly alters the timing of the pandemic, but does little to mitigate the economic and human impacts: apart from a delay in the main phase of the pandemic, most variables behave similar to the \textit{No Intervention} scenario.

This indifference between the intensities of the interventions is likely related to the relationship we identified between the GDP shortfall and the infection rate reduction, as shown in Figure \ref{fig:prodinfectscatter}. There are strong diminishing returns, such that even an intervention of a low intensity (5\%) already reduces the infection rate substantially (60\%), and that additional measures have a lower effect on the infection rate.

Essentially, the intensity of the intervention -- above a certain minimum level -- appears to be less important than the timing of the intervention. Again, this finding contradicts the study of \cite{guan_global_2020}, our results indicating that intervention should perhaps be \emph{less} strict, as the intervention intensity faces strong diminishing returns.

\subsection{For How Long to Intervene?}
\begin{figure}
    \centering
    \includegraphics[width=\textwidth]{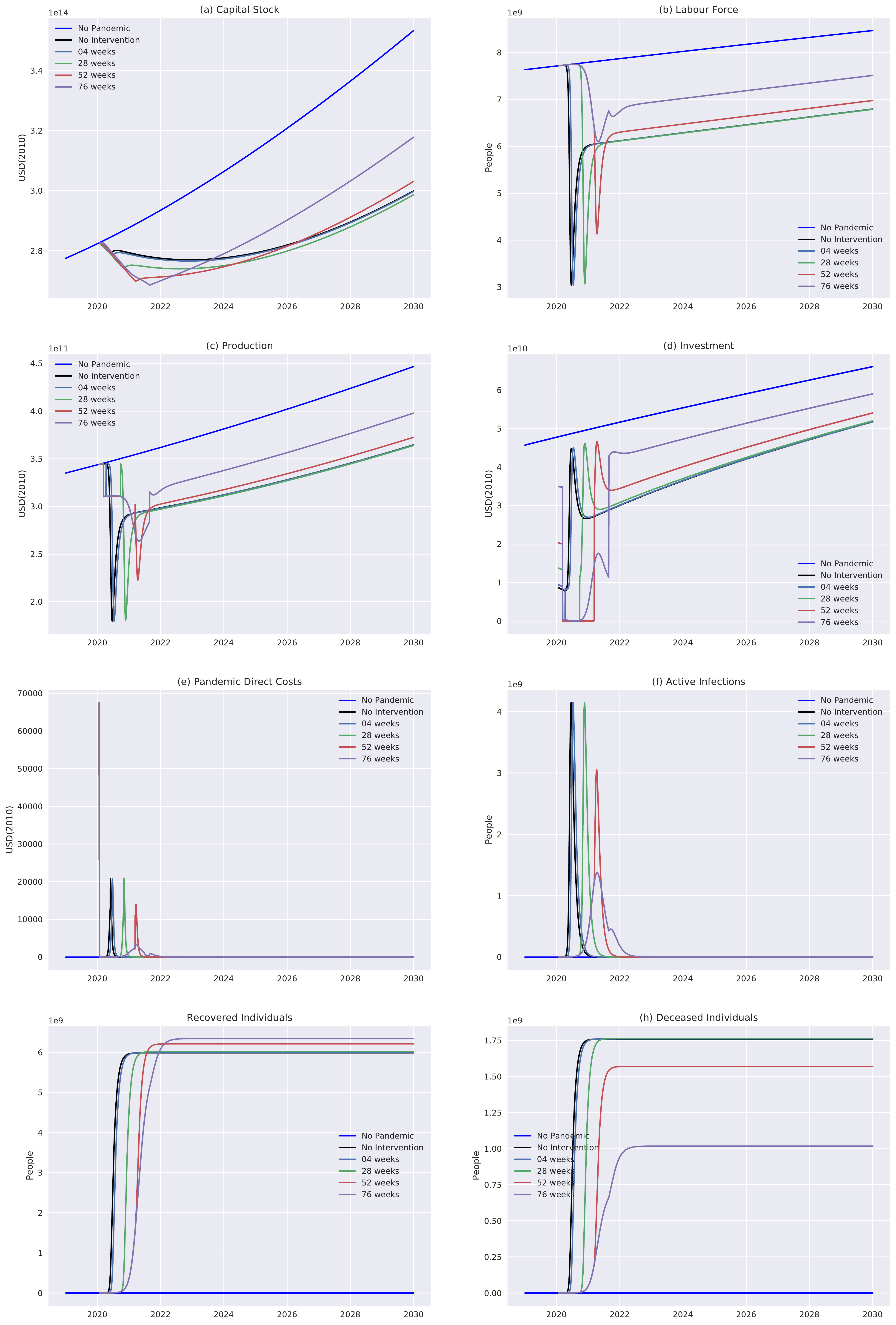}
    \caption{Simulation results when varying the duration of the policy intervention.}
    \label{fig:durations}
\end{figure}

To analyse the impact of the duration of the policy intervention, we vary the duration of the policy intervention, whilst maintaining the intervention intensity fixed at 10\% and the starting date fixed at March 12. Figure \ref{fig:durations} shows the results of simulating intervention durations of 4 weeks, 28 weeks, 52 weeks and 76 weeks. It is clear from the figure that the duration of the policy intervention can have a large impact on the trajectory of the pandemic, and its human and economic aftermath.

From Figure \ref{fig:durations}, it seems that policies with longer durations clearly lead to lower human and economic impacts, with a 76-week duration -- the longest of our simulations -- showing dramatically lower number of total mortalities, as well as a much quicker post-pandemic recovery of production. The key appears to be that 76 weeks is sufficient to include the peak in active infections of the trajectory determined by the reduced infection rate. This observation suggests that policies with a lower intensity would require a shorter duration, whereas policies with a higher intensity would require a longer duration -- something which may, at first, appear counter-intuitive. 

Our results partially support the findings of \cite{eichenbaum_macroeconomics_2020}, who warn of ending the intervention ``too early'', but contradict the claims of \cite{guan_global_2020} that intervention should be ``short'' -- although the duration of the intervention should naturally be as short as possible. Our results suggest that stricter policies will require longer durations.

\subsection{Considerations, Limitations, and Concerns}
We have tried to make sensible modelling choices in this study, but, like all models, our model is a simplification that focuses only on certain aspects and ignores others. The simulation results should not be understood \emph{literally}: the intention of our model has never been to provide numerically accurate predictions, but to generate insights into the impacts of policy interventions by analysing the dynamics of the system as a whole. Although the insights from the numerical simulations can contribute to improving policy interventions during pandemics, it is important to appreciate the limitations of our model and results.

We are concerned about the quality of the parameters used for the epidemiological part of the model: there is a great deal of doubt and uncertainty about the quality of the official datasets on the spread of the pandemic -- that is, the number of confirmed cases, recovered cases and mortalities. There is a general sense that the number of confirmed cases is not representative for the number of infections, as testing is severely lacking in many regions. The lack of testing also affects the number of mortalities due to the pandemic, and number of recovered cases. Although we have used the data that is available without much discrimination, we share the concerns of many other researchers as to the quality of this data.

It is also unusual for a model of economic growth to operate at a daily resolution. We do not think this directly invalidates our resulting insights, although it means that the parameter values may appear unusual to researchers and practicioners, and that special care must be taken in the interpretation of the results. An alternative would be to develop the model in continuous time, which might be more familiar to some. However, in that case it would be necessary to discretise the model later for performing numerical experiments -- the model would, in the end, be the same, so presenting the model directly in discrete time appears to be a simpler alternative.

The parameter values were chosen for the model to represent the global economy and the global spread of COVID-19. There are, however, large differences between regions in the world. For instance, the five countries used for estimating the economic impact of policy measures -- France, Germany, Italy, Spain, and the United Kingdom, which were chosen for their data availability -- are probably not entirely representative for the rest of the world. There is also no global central government that implements global policy, and our insights are therefore not directly applicable by any specific authority. The purpose of the study, however, was not to generate recommendations for specific actions, but to generate insights into the impacts and trade-offs that policy interventions must consider. Regional, national and local policy can differ from ``global'' policy -- and probably should, as policy can be optimised to local conditions -- but the insights on intervention timing, intensity and duration may nevertheless be useful at these levels also. It would be possible to adapt the model for use at regional, national or local scales. In this case, we would recommend considering replacing the linear admission costs with a specification that allows for increasing marginal costs of admissions, which might better reflect increasing costs in the short run due to capacity saturation.

For calibrating the model, we estimated the economic impacts using an estimate of the shortfall in electricity consumption. Although we believe this approach is valid, it is difficult to assess the accuracy of the economic impact estimates. The approach based on electricity consumption could also be complemented by other near-real-time, high-resolution data sources that are believed to correspond well to economic activity, such as satellite observations, mobile phone movement and activity data, and urban traffic data. However, these data are not usually as widely available as electricity consumption data.

The model does not incorporate any demographic heterogeneity. Since some pandemics appear to affect people with certain demographic characteristics differently, this may bias the results. For instance, the mortality rate of COVID-19 appears to differ greatly between old and young people: if the disease has a greater impact on groups that were not originally included in the work force anyway, the model could exaggerate the economic impact of the pandemic by disconsidering demographic heterogeneity. We do not believe this to affect the main insights derived from our model simulations, since we do not think it substantially alters the dynamics of the system. However, it would certainly be an issue for the ``predictive accuracy'' of the simulations.

Since the model is deterministic, agents in the model have perfect foresight from the very start of the simulation. This is, naturally, not true in the real world, in which there are large uncertainties about future developments. This gives the model agents an unrealistic ability to plan for the future, and the economic portion of the model should therefore be considered an optimistic path. Another detail that also may positively bias the outcomes, is that the model does not include structural damages -- such as bankruptcies, institutional change, changing habits, and so forth -- and affords the model agents much more flexibility than economic agents may have in reality, where they may be facing additional restrictions.

Finally, we only simulated very simple policies for the purpose of understanding the impact of altering the policy in a very specific way. For instance, superior policies can be made relatively easily by allowing the intensity of the intervention to vary during the pandemic. Our examples, in which starting dates, intensity and duration are fixed, only served for illutration and to understand some of the dimensions of policy intervention.

We reiterate that our purpose has not been to provide numerically accurate predictions -- nor the means to generate accurate predictions -- of the evolution of the COVID-19 pandemic or the global economy. We have only explored particular aspects of effective policy responses to a pandemic, using a very high-level and theoretical approach, and it is with this in mind that our results are most appropriately appreciated. Our research does not aim to offer specific guidance for world authorities on the handling of the COVID-19 pandemic, but to analyse how a pandemic interacts with the global economy and thus help establish a set of of general guidelines.

\section{Conclusion}
We have presented a mathematical model for the joint evolution of the economy and a pandemic, based on incorporating the dynamics of the SIR model that describes the spread of epidemics into a neoclassical economic growth model framework. This model is subsequently adapted to represent the global economy under the impact of the COVID-19 pandemic by selecting appropriate functional forms and parameter values. The model includes a parameter that represents policy, by which economic production can be lowered in exchange for a reduced infection rate.

Using the calibrated model, we simulate the joint evolution of the economy and the pandemic for a series of policy assumptions in order to discover what is the most effective timing of a policy intervention, what intensity of policy intervention is most effective, and how long policy intervention should last.

Our experiments suggest that it is most effective to start the policy intervention slightly before the number of confirmed cases grows at its highest rate -- that is, to wait until the disease is sufficiently spread. Not only does this help lower the peak in active infections, it also reduces the economic impact and the number of mortalities. Starting too early can delay the pandemic, but does not otherwise significantly alter its course, whereas starting after the peak in active infections can obviously not impact the peak.

Furthermore, altering the intensity of the intervention does not appear to greatly influence the evolution of the pandemic nor the economy, other than cause minor delays. We ascribe the lack of effect to the concave relationship that we estimated between intervention intensity and infection rate reduction, as a large reduction in infection rate can be achieved by sacrificing a modest proportion of economic production, and appears to show strong decreasing returns thereafter. Our estimates suggest that a 60\% reduction in the infection rate can be achieved by sacrificing only 5\% of production, whereas a 70\% reduction in infection rate could be achieved for a 10\% reduction in production.

Altering the duration of the intervention showed that interventions with a longer duration lead to significantly lower mortalities and a quicker post-pandemic recovery in economic production. The key observation is that the policy must include the peak of the \textit{new} path set out by the reduced infection rate: in short, policy intervention should last until the peak has passed. Therefore -- somewhat counter-intuitively -- stricter policies should last longer, and less strict policies should last shorter.

Although the scenarios we present are not necessarily numerically accurate as predictions -- mostly due to generalisations made for modelling purposes, large regional variations, and large uncertainties in the parameters -- our conclusions are based mainly on the dynamics revealed by the policy experiments, and not specifically on their numerical values. As such, we hope that our model can serve as a tool for enhancing our understanding of the design of effective policies against the spread of pandemics, and that our insights can contribute to this discussion and provide general guidelines for policymakers.

\bibliographystyle{elsarticle-harv}
\bibliography{PPGEAPandemicsTaskForce}

\section*{Data and Code Availability}
All the data used in this study is available to the public, and the various data sources have been referenced at the appropriate places along the study.

The custom computer code for running the simulations is available at \url{https://github.com/iantrotter/ME3PI-SIM}.

\end{document}